\documentclass{revtex4}
\usepackage{makeidx}
\usepackage{amssymb}
\usepackage{amsmath}
\usepackage{mathrsfs}
\usepackage{graphicx}
\usepackage{dcolumn}
\usepackage{bm}
\usepackage[center]{subfigure}
\usepackage{color}

\begin{document}

\title{Spin-orbit coupling and nonlinear modes of the polariton condensate
in a harmonic trap}
\author{Hidetsugu Sakaguchi$^{1}$, Boris A. Malomed$^{2,3,6}$, and Dmitry V.
Skryabin$^{4,5}$}
\address{$^{1}$Department of Applied Science for Electronics and Materials,
Interdisciplinary Graduate School of Engineering Sciences, Kyushu
University, Fukuoka, Japan \\
$^{2}$Department of Physical Electronics, School of Electrical Engineering,
Faculty of Engineering, Tel Aviv University, Tel Aviv, Israel\\
$^3$Laboratory of Nonlinear-Optical Informatics, ITMO University, St. Petersburg 197101, Russia\\
$^{4}$Department of Physics, University of Bath, Bath, UK\\
$^5$Department of Nanophotonics and Metamaterials, ITMO University,  St. Petersburg 197101, Russia\\
$^6$Author to whom any correspondence should be addressed.\\
{\bf Email: malomed@post.tau.ac.il}}

\begin{abstract}
We consider a model of the exciton-polariton condensate based on a system of
two Gross-Pitaevskii equations coupled by the second-order differential
operator, which represents the spin-orbit coupling (SOC) in the system. Also
included are the linear gain, effective diffusion, nonlinear loss, and the
standard harmonic-oscillator trapping potential, as well as the Zeeman
splitting. By means of combined analytical and numerical methods, we
identify stable two-dimensional modes supported by the nonlinear system. In
the absence of the Zeeman splitting, these are mixed modes, which combine
zero and nonzero vorticities in each of the two spinor components, and
vortex-antivortex complexes. We have also found a range of parameters where
the mixed-mode and vortex-antivortex states coexist and are stable.
Sufficiently strong Zeeman splitting creates stable semi-vortex states, with
vorticities $0$ in one component and $2$ in the other.
\end{abstract}

\maketitle

\section{Introduction}

A rapidly advancing direction in the studies of multidimensional nonlinear
wave patterns deals with two-component systems supporting the spin-orbit
coupling (SOC), which are represented by derivative linear-mixing terms in
the underlying systems of nonlinear Schr\"{o}dinger/Gross-Pitaevskii
equations (GPEs) \cite{dal,gal,zha}. This topic has drawn a great deal of
interest after it has been demonstrated, both theoretically and
experimentally, that the SOC of nonlinear spinor fields, which has been
known for a long time as a fundamentally important interaction in the
quantum mechanics of electrons in semiconductor settings, can be emulated in
a variety of bosonic media, such as atomic Bose-Einstein condensates (BECs)
\cite{dal,gal,zha}, polariton microcavity systems \cite{she,sal}, and
photonics \cite{bli}.

Recent theoretical analysis of systems combining the cubic attractive
interactions of the fields and linear SOC has produced several unexpected
results. First, in the free-space two-dimensional (2D) geometry, the SOC
governed by the Rashba Hamiltonian for BEC\ gives rise to \emph{stable}
half-fundamental-half-vortex solitons of two types, \textit{viz}., \textit{%
semi-vortices}, with one fundamental (zero-vorticity) component, and one
with vorticity $S=\pm 1$, and \textit{mixed modes}, which combine
fundamental and vortex terms with $S=\pm 1$ in both field components \cite%
{sak}. Stability of nonlinear states in SOC systems is radically different
with respect to the ones without SOC \cite{skr,HPu}. Indeed, the GPE in the
free 2D space with the cubic self-attraction, or a system of two GPEs with
self- and cross attraction, gives rise to families of zero-vorticity, alias
fundamental (Townes' \cite{Townes}) and vorticity-carrying (\cite{Yank}-\cite%
{Minsk2}) bright solitons which are completely unstable either due to the
critical collapse \cite{coll1,coll2,coll3} or due to the ring-splitting
instabilities \cite{firth1,firth2}, respectively. On the other hand, the SOC
terms come with a coefficient which fixes an inverse length scale in the
system, thus breaking the scaling invariance which makes norms of all the
solitons belonging to the Townes' family exactly equal to the critical value
necessary for the onset of the collapse. As a result, norms of the
two-component solitons of both the semi-vortex and mixed-mode types, morphed
by the interplay of the cubic attraction and SOC, are pushed \emph{below}
the collapse threshold, which immediately makes them stable \cite%
{short-review}. Actually, the stabilized 2D solitons play the role of the
ground state, which was missing in the system with the cubic attraction in
the absence of SOC. Furthermore, although in the 3D free-space binary BEC
the SOC terms, unlike the 2D case, cannot create the missing ground state,
as the corresponding \textit{supercritical collapse} has zero threshold \cite%
{coll1,coll2,coll3}, the linear SOC terms readily create metastable solitons
of the same two types, semi-vortices  and mixed modes \cite{HPu}.

In microcavity polariton systems, the largest contribution to SOC originates
from the energy splitting between the TE and TM modes. If written in the
spinor form, it is tantamount to the linear inter-component coupling
represented by the second-order spatial derivatives \cite{fla}, while the
above-mentioned Rashba SOC in BEC is accounted for by the first-order
derivatives, in the leading approximation \cite{zha,HPu}. Thus, depending on
physics of the SOC system under the consideration, vorticities in the two
spinor components differ by either $\Delta S=2$ or $1$, for the second- and
first-order-derivative coupling terms, respectively.

Adding trapping or lattice potentials to the nonlinear SOC systems is a very
active direction of the current research, see, e.g., Refs. \cite%
{kar,sch,kart}. In the polariton microcavities, which are the subject of the
present work, a variety of practical methods have been developed to secure
high quality of the engineered trapping landscapes \cite{sch}. As concerns
polaritons in a simple single potential well, there are only a few studies
that addressed nonlinear modes of these traps, largely disregarding SOC,
see, e.g., Refs. \cite{ask,tra,yul,bor1}. Nevertheless, a recent
experimental work on open polaritonic resonators supplies a system with
strong SOC effects \cite{duf}.

In this work, we address an effectively 2D polariton condensate modeled by a
system of two GPEs with the SOC terms represented by the above-mentioned
second spatial derivatives. Usual ingredients of microcavity models, \textit{%
viz}., the linear gain, viscosity (diffusion of the effective fields),
nonlinear cubic loss, and an isotropic harmonic-oscillator trapping
potential, are included. In the general case, the Zeeman splitting between
the two fields is present too. Our objective is to identify stable 2D states
in this system, which turn out to be mixed modes and vortex-antivortex bound
states, that tend to be stable, severally, under the action of weak and
strong SOC, with a small bistability area in the parameter space. These
results for the symmetric system, which does not include Zeeman splitting,
are obtained by means of analytical and numerical methods in Section II. In
addition to that, in Section III we demonstrate that sufficiently strong
Zeeman splitting creates stable semi-vortex modes, with vorticities $0$ and $%
2$ in its components. The latter states are often referred to as the
half-vortices in the polariton context \cite{fla,rub,nat}, and are also
known in models without SOC \cite{rub}, or with the spin-only coupling
(direct linear mixing of the Rabi-coupling type) \cite{bor1}; see also Ref.
\cite{bor2}, as concerns vortex lattices in a scalar model, which does not
include spin effects. The paper is completed by a summary in Section IV.

\section{Analytical and numerical results for the symmetric system}

\subsection{Basic equations}

The model that we consider in this work consists of coupled GPEs for the
two-component polariton spinor wave function $(\Psi _{+},\Psi _{-})$ \cite%
{fla,ask}. We assume that the polariton condensate is pumped by means of a
nonresonant optical scheme, which creates a density of coherent and
incoherent excitons \cite{wou,kee}. Adiabatic elimination of the latter
yields the following system of coupled equations \cite{kee}:
\begin{gather}
i\partial _{t}\Psi _{+}=-{\frac{1}{2}}(1-i\eta )(\partial _{x}^{2}+\partial
_{y}^{2})\Psi _{+}+(|\Psi _{+}|^{2}+\alpha |\Psi _{-}|^{2})\Psi _{+}  \notag
\\
+\beta (\partial _{x}-i\partial _{y})^{2}\Psi _{-}+i(\varepsilon -\sigma
|\Psi _{+}|^{2})\Psi _{+}+\left[ \frac{1}{2}\left( x^{2}+y^{2}\right)
+\Omega \right] \Psi _{+},  \label{+} \\
i\partial _{t}\Psi _{-}=-{\frac{1}{2}}(1-i\eta )(\partial _{x}^{2}+\partial
_{y}^{2})\Psi _{-}+(|\Psi _{-}|^{2}+\alpha |\Psi _{+}|^{2})\Psi _{-}  \notag
\\
+\beta (\partial _{x}+i\partial _{y})^{2}\Psi _{+}+i(\varepsilon -\sigma
|\Psi _{-}|^{2})\Psi _{-}+\left[ \frac{1}{2}\left( x^{2}+y^{2}\right)
-\Omega \right] \Psi _{-}.  \label{-}
\end{gather}%
Here, gain $\varepsilon >0$ is determined by the rate of the pumping of the
polariton density, $\sigma >0$ is the gain-saturation coefficient that
prevents an unphysical blow-up and allows for the time-independent finite
density solutions to exist in the system, and $\eta $ is an effective
polariton diffusion coefficient (viscosity), linked to the carrier diffusion
\cite{kee,sie}. Further, the strength of the repulsive intra-component
interactions is normalized to be $1$, while $\alpha =-0.05$ accounts for the
inter-component attraction, and $\beta $ is the strength of the polariton
SOC \cite{fla}, while the strength of the harmonic-oscillator potential is
scaled to be $1$. The potential corresponds to the shift of the cavity
detuning, and originates from the trapping of the photonic component of
exciton-polariton condensate. The potential can be created, for example, if
either the above-mentioned open cavity with a parabolic top mirror is used,
or one of the many available patterning techniques is applied to the top
mirror in the traditional closed-resonator setting with Bragg reflectors
\cite{sch}. Lastly, $\Omega $ characterizes the Zeeman-splitting energy,
which is proportional to the applied magnetic field. Accessible physical
values of the parameters in the linear Hamiltonian part of Eqs. (\ref{+}), (%
\ref{-}), and, in particular, the SOC strength in the presence of the
harmonic-oscillator potential have been discussed in detail in Ref. \cite%
{duf}. That work demonstrates that SOC energy may reach $1$meV, which is
comparable to the characteristic harmonic-oscillator energy, thereby making
values of $\beta $ between $0$ up to $1$ physically relevant. Achieving
large polarization energy splitting has been important, since the beginning
of the studies of microcavity polaritons, in the context of the polariton
spin Hall effect \cite{kav,ley}, see also Ref. \cite{she} for a detailed
review of the previous work. Currently, the design of strong SOC is getting
crucially important for the work towards experimental observation and
applications of polariton topological insulators \cite{sal,bar,nal,kar1}.

Stationary solutions to Eqs. (\ref{+}) and (\ref{-}), with chemical
potential $\mu $, are sought in the form of
\begin{equation}
\Psi _{\pm }=e^{-i\mu t}\psi _{\pm }\left( r,\theta \right) ,  \label{mu}
\end{equation}%
where $\psi _{\pm }$ are complex functions of polar coordinates $\left(
r,\theta \right) $. The substitution of expressions (\ref{mu}) in Eqs. (\ref%
{+}) and (\ref{-}) leads to the following stationary equations:%
\begin{eqnarray}
&&\mu \psi _{+}=-{\frac{1}{2}}(1-i\eta )\left( \frac{\partial ^{2}}{\partial
r^{2}}+\frac{1}{r}\frac{\partial }{\partial r}-\frac{1}{r^{2}}\frac{\partial
^{2}}{\partial \theta ^{2}}\right) \psi _{+}+(|\psi _{+}|^{2}+\alpha |\psi
_{-}|^{2})\psi _{+}  \notag \\
&&+\beta \left[ e^{-i\theta }\left( \frac{\partial }{\partial r}-\frac{i}{r}%
\frac{\partial }{\partial \theta }\right) \right] ^{2}\psi
_{-}+i(\varepsilon -\sigma |\psi _{+}|^{2})\psi _{+}+\left( \frac{1}{2}%
r^{2}+\Omega \right) \psi _{+},  \label{u+} \\
&&\mu \psi _{-}=-{\frac{1}{2}}(1-i\eta )\left( \frac{\partial ^{2}}{\partial
r^{2}}+\frac{1}{r}\frac{\partial }{\partial r}-\frac{1}{r^{2}}\frac{\partial
^{2}}{\partial \theta ^{2}}\right) \psi _{-}+(|\psi _{-}|^{2}+\alpha |\psi
_{+}|^{2})\psi _{-}  \notag \\
&&+\beta \left[ e^{i\theta }\left( \frac{\partial }{\partial r}+\frac{i}{r}%
\frac{\partial }{\partial \theta }\right) \right] ^{2}\psi
_{+}+i(\varepsilon -\sigma |\psi _{-}|^{2})\psi _{-}+\left( \frac{1}{2}%
r^{2}+\Omega \right) \psi _{-},  \label{u-}
\end{eqnarray}%
which are used below as the basic system.

\subsection{Linear and quasi-linear modes: vortex-antivortex states}

Before tackling the full nonlinear system, it is relevant to address the
simplest version of stationary equations (\ref{u+}) and (\ref{u-}), which
neglects the nonlinearity, gain, and diffusion, as well as assumes the
symmetry between the components, i.e., $\Omega =0$ (no Zeeman splitting):
\begin{gather}
\mu \psi _{+}=-{\frac{1}{2}}\left( \frac{\partial ^{2}}{\partial r^{2}}+%
\frac{1}{r}\frac{\partial }{\partial r}-\frac{1}{r^{2}}\frac{\partial ^{2}}{%
\partial \theta ^{2}}\right) \psi _{+}+\beta \left[ e^{-i\theta }\left(
\frac{\partial }{\partial r}-\frac{i}{r}\frac{\partial }{\partial \theta }%
\right) \right] ^{2}\psi _{-}+\frac{1}{2}r^{2}\psi _{+},  \notag \\
\mu \psi _{-}=-{\frac{1}{2}}\left( \frac{\partial ^{2}}{\partial r^{2}}+%
\frac{1}{r}\frac{\partial }{\partial r}-\frac{1}{r^{2}}\frac{\partial ^{2}}{%
\partial \theta ^{2}}\right) \psi _{-}+\beta \left[ e^{i\theta }\left( \frac{%
\partial }{\partial r}+\frac{i}{r}\frac{\partial }{\partial \theta }\right) %
\right] ^{2}\psi _{+}+\frac{1}{2}r^{2}\psi _{-}.  \label{lin}
\end{gather}

It is easy to see that Eq. (\ref{lin}) gives rise to two exact eigenmodes of
the linearized dissipation-free system, corresponding to independent signs $%
\pm $ in Eq. (\ref{vortex-antivortex}):%
\begin{equation}
\psi _{+}=A_{\max }r\exp \left( -i\theta -\frac{r^{2}}{2\sqrt{1\mp 2\beta }}%
\right) ,~\psi _{-}=\pm A_{\max }r\exp \left( i\theta -\frac{r^{2}}{2\sqrt{%
1\mp 2\beta }}\right) ,  \label{vortex-antivortex}
\end{equation}%
\begin{equation}
\mu _{\pm }^{(0)}=2\sqrt{1\mp 2\beta },  \label{mu0}
\end{equation}%
where $A_{\max }$ is an arbitrary constant, and superscript $(0)$ attached
to $\mu _{\pm }$ refers to the linear approximation. These exact eigenmodes,
which, as a matter of fact, are found following the pattern of standard
solutions for the 2D isotropic harmonic oscillator in quantum mechanics, may
be naturally called vortex-antivortex states. Note that the symmetric
vortex-antivortex state with the lower eigenvalue of the energy, $\mu
_{+}^{(0)}=2\sqrt{1-2\beta }$, exists if the SOC is not too strong, \textit{%
viz}., at $\beta \leq 0.5$, while its antisymmetric counterpart with higher
energy, $\mu _{-}^{(0)}=2\sqrt{1+2\beta }$, exists at all values of $\beta $%
. In fact, the disappearance of the eigenmode (\ref{vortex-antivortex})
corresponding to the top sign in $\mp $ at $\beta =0.5$, where the radial
size of the eigenmode shrinks to zero as $\left( 1-2\beta \right) ^{1/4}$,
may be considered as a quantum phase transition in the polariton condensate,
cf. Ref. \cite{phase-transition}.

Both types of the vortex-antivortex states, antisymmetric and symmetric
ones, were experimentally observed, in a quasi-linear regime, in Ref. \cite%
{duf}, where they were named, respectively, modes of types (i) and (iii). In
terms of the present notation, the respective SOC\ coupling, estimated in
Ref. \cite{duf}, was $\beta \approx 0.04$.

In a sense, the exact vortex-antivortex modes are similar to the
above-mentioned semi-vortices  found in the nonlinear dissipation-free 2D\
SOC\ model of the atomic condensate considered in Ref. \cite{sak}, which
feature zero vorticity in one component and vorticity $\pm 1$ in the other.
Indeed, the difference of the vorticities in the components of the
semi-vortex is $\Delta S=1$ because (as mentioned above too) SOC in the
atomic condensates is represented by the linear differential operator of the
first order \cite{zha}, while in the present system the difference between
the components of the vortex-antivortex is $\Delta S=2$, as the SOC is
represented by the second-order operator in Eqs. (\ref{+}) and (\ref{-}).

If the nondissipative nonlinear terms are kept in Eqs. (\ref{u+}) and (\ref%
{u-}) as small perturbations, it is easy to find the first nonlinear
correction to the chemical potential, $\mu _{\pm }^{(1)}$, using the
commonly known method in quantum mechanics, that generates the first
correction as the spatial average of the perturbation potential \cite{skr,LL}%
. A simple calculation, which makes use of the unperturbed wave functions (%
\ref{vortex-antivortex}), yields

\begin{equation}
\mu _{\pm }^{(1)}=\frac{1+\alpha }{4}\sqrt{1\mp 2\beta }A_{\max }^{2},
\label{mu1}
\end{equation}%
where $A_{\max }^{2}$ should be small enough, to keep correction (\ref{mu1})
small in comparison with the result (\ref{mu0}), obtained in the linear
limit.

Furthermore, if the linear gain, diffusion, and nonlinear loss are also kept
as small perturbations in Eqs. (\ref{u+}) and (\ref{u-}), then amplitude $%
A_{\max }$ of the linear mode (\ref{vortex-antivortex}), which is treated in
Eqs. (\ref{vortex-antivortex}) and (\ref{mu1}) as a free parameter, is
uniquely selected by the obvious balance condition for the total norm (it is
defined below by Eq. (\ref{N})), written as%
\begin{equation}
\varepsilon \int_{0}^{\infty }\left\vert \psi _{\pm }(r)\right\vert
^{2}rdr=\sigma \int_{0}^{\infty }\left\vert \psi _{\pm }(r)\right\vert
^{4}rdr+\frac{\eta }{2}\int_{0}^{\infty }\left[ \left\vert \frac{d\psi _{\pm
}}{dr}\right\vert ^{2}+\frac{\left\vert \psi _{\pm }(r)\right\vert ^{2}}{%
r^{2}}\right] rdr,  \label{balance}
\end{equation}%
for the mode $\left\vert \psi _{+}(r)\right\vert =|\psi _{-}(r)|$ taken as
per Eq. (\ref{vortex-antivortex}). The result of the simple calculation is%
\begin{equation}
\left( A_{\max }\right) _{\mathrm{balance}}^{2}=\frac{\varepsilon \sqrt{1\mp
2\beta }}{2\sigma \left( 1\mp 2\beta \right) +\eta },  \label{A^2}
\end{equation}%
which makes sense if the pump rate $\varepsilon $ is small enough. A more
sophisticated analytical approximation for the vortex-antivortex in the full
system is developed below, see Eqs. (\ref{ans})-(\ref{NN}).

\subsection{Mixed-mode states}

\begin{figure}[tbp]
\begin{center}
\includegraphics[height=10cm]{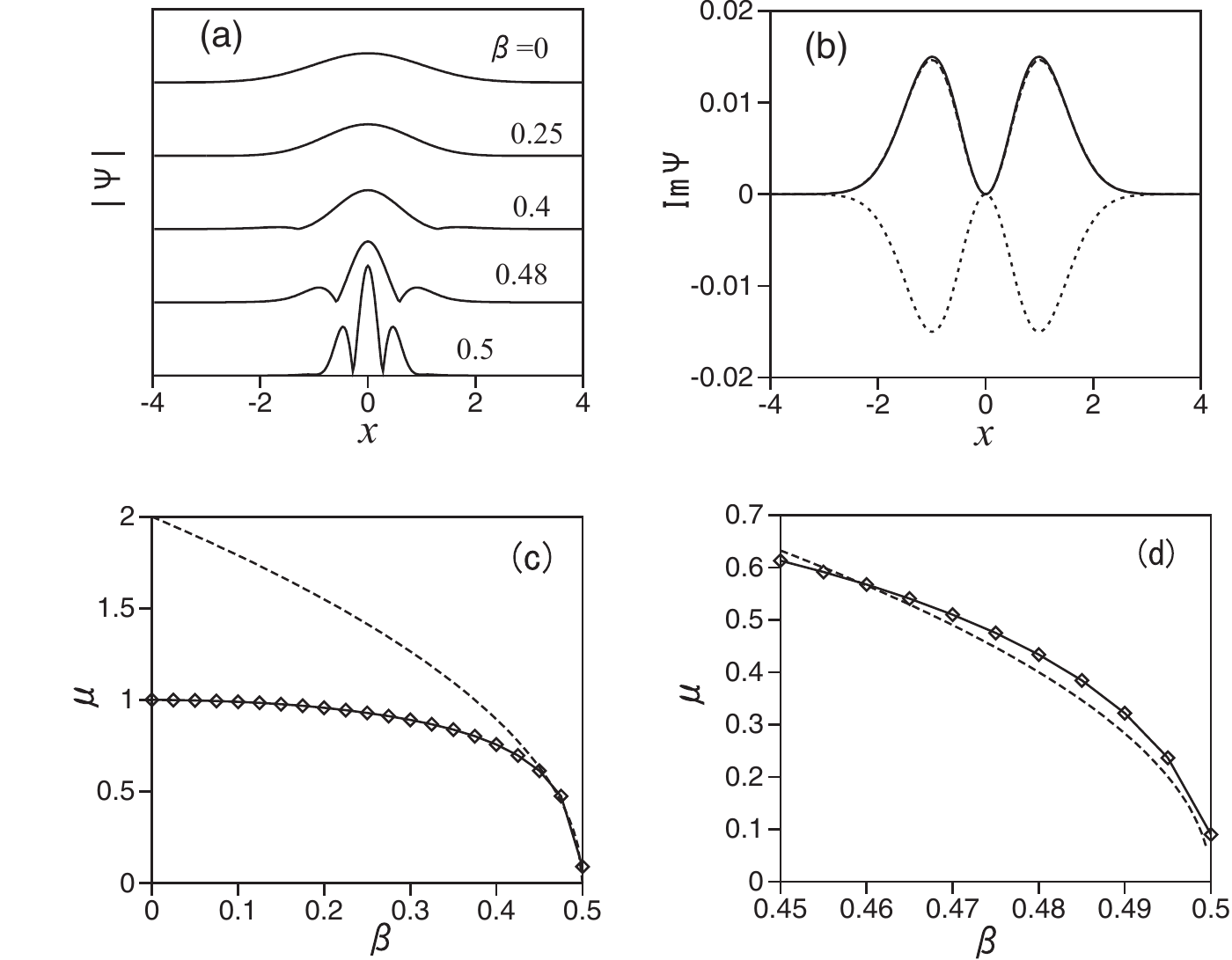}
\end{center}
\caption{(a) Profiles of $|\protect\psi _{+}|=|\protect\psi _{-}|$ for
mixed-mode eigenstates of the linearized system (\protect\ref{lin}),
obtained as numerical imaginary-time solutions for values of the SOC
strength $\protect\beta =0.5,0.48,0.4$, $0.35$ and $0$ (from bottom to top).
(b) Imaginary parts of the numerically found solution for $\protect\psi _{+}$
(solid curve), and $\protect\psi _{-}$ (dotted curve), shown along the
diagonal cross section, $y=x$ (i.e., $\protect\theta =\protect\pi /4$), for $%
\protect\beta =0.1$. The dashed line, which completely overlaps with the
solid one, is the analytical prediction, Im$\left( \protect\psi _{+}\right)
=2A_{\max }\protect\beta x^{2}\exp \left( -x^{2}\right) $, given by Eq. (%
\protect\ref{mixed mode}). (c) Energy eigenvalues of the mixed-mode and
symmetric vortex-antivortex states, shown by the solid and dashed lines,
respectively, versus the SOC strength, $\protect\beta $. The exact
vortex-antivortex' eigenvalue is given by Eq. (\protect\ref{mu0}) with the
upper sign. (d) Energy eigenvalues of the mixed-mode and vortex-antivortex
states for $0.45<\protect\beta <0.5$.}
\label{f1}
\end{figure}

In addition to the two species of vortex-antivortex complexes, built as per
Eqs. (\ref{vortex-antivortex}), the linearized coupled GPEs (\ref{lin}) give
rise to another type of eigenstates, which, following Ref. \cite{sak}, may
be called mixed modes, as they contain terms with zero vorticity and
vorticities $\pm 2$ in each component. Like the symmetric vortex-antivortex
states (\ref{vortex-antivortex}), the mixed modes obey constraint $\psi
_{+}=\psi _{-}^{\ast }$, with $\ast $ standing for the complex conjugate.
Linear mixed modes cannot be represented by an exact solution (unlike Eq. (%
\ref{vortex-antivortex}) for the vortex-antivortex complexes), but they can
be easily constructed in an approximate form for $\beta \ll 1$, starting
from an obvious ground state of the harmonic-oscillator in each component at
$\beta =0$:
\begin{equation}
\psi _{\pm }\approx A_{\max }\exp \left( -\frac{r^{2}}{2}\right) \left[ 1\pm
i\beta r^{2}\sin (2\theta )\right] ,\;\mu \approx 1-\beta .
\label{mixed mode}
\end{equation}%
At larger $\beta $, the mixed mode eigenmodes were constructed numerically
by means of the imaginary-time method, applied to the time-dependent version
of linearized equations (\ref{lin}), starting with the ground-state input, $%
\Psi _{+}=\Psi _{-}=\exp (-r^{2}/2)$. The numerically found mixed-mode
shapes are displayed in Fig. \ref{f1}, and the comparison with the
analytical approximation (\ref{mixed mode}), for a relevant small value $%
\beta =0.1$, is presented in Fig. \ref{f1}(b).

Further, eigenvalues corresponding to the mixed modes were compared to the
analytically found eigenvalues for the vortex-antivortex states, given by
Eq. (\ref{mu0}), with the aim to identify the ground state, which must have
the smallest eigenvalue of energy, $\mu $, for given $\beta $. As a result,
it was found that, at $\beta <0.46$, the mixed modes realize the ground
state (for small $\beta $, this is evident from the comparison of Eqs. (\ref%
{mu0}) and (\ref{mixed mode})), but in a narrow interval of $0.46<\beta <0.5$%
, the symmetric vortex-antivortex complex, with the upper sign in Eq. (\ref%
{vortex-antivortex}), produces a slightly smaller value of $\mu $, which is
given by Eq. (\ref{mu0}), as shown in Fig. \ref{f1}(c); recall that
vortex-antivortex state with the upper sign does not exist at $\mu >0.5$.
The same figure \ref{f1}(c) shows that the mixed mode does not exists either
at $\beta >0.5$ (which is another essential manifestation of the
above-mentioned quantum phase transition), hence the only eigenmode which
survives at $\beta >0.5$ is the antisymmetric vortex-antivortex complex with
the lower sign in Eq. (\ref{vortex-antivortex}). It is worthy to mention
here that, as shown below, solely vortex-antivortex modes with \emph{%
opposite signs} of their components are found as stable states in the full
system of Eqs. (\ref{+}) and (\ref{-}), while vortex-antivortex bound states
with identical signs of its components cannot be generated by the full
system, even if they may realize the ground state of the linearized system
at $\beta $ close to $0.5$.
\begin{figure}[tbp]
\begin{center}
\includegraphics[height=5cm]{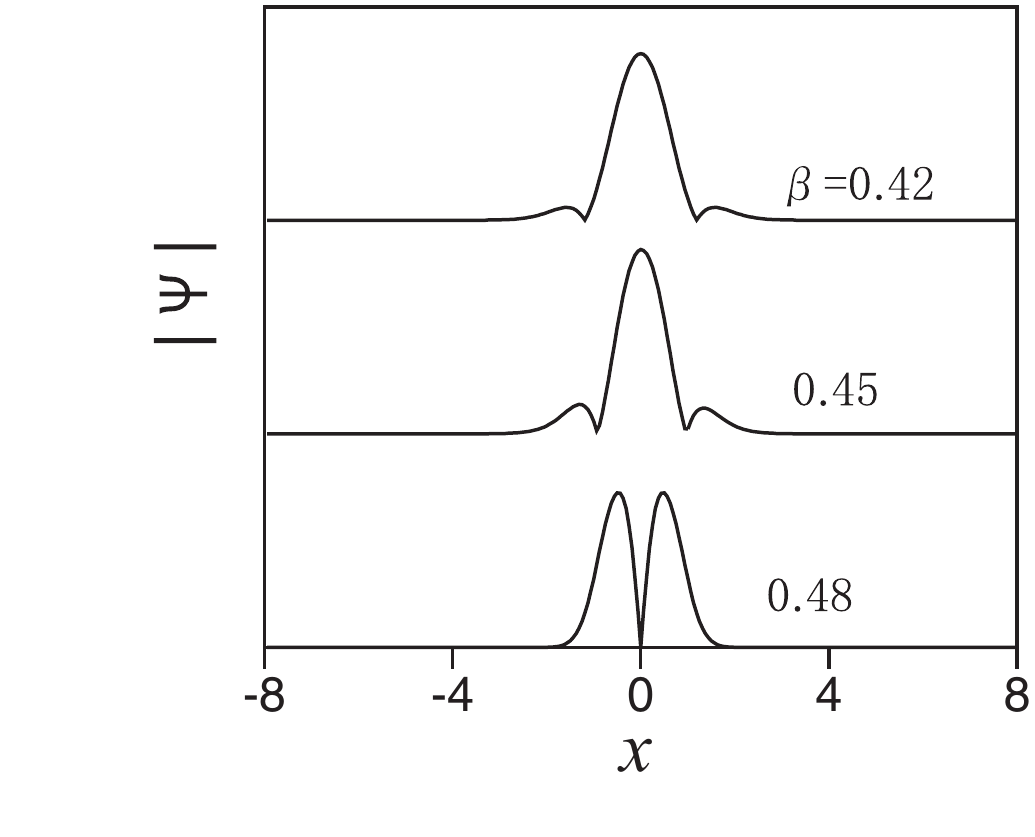}
\end{center}
\caption{A set of profiles of components $A\equiv |\protect\psi _{+}|=|%
\protect\psi _{-}|$ of the ground state of the nonlinear dissipation-free
system (\protect\ref{nonlin}) at values of the SOC$\ $strength $\protect%
\beta =0.48,0.45,0.42$ (from bottom to top). A transition from the mixed
mode to the symmetric vortex-antivortex complex takes place as $\protect%
\beta $ increases from $0.45$ to $0.48$.}
\label{f2}
\end{figure}
\begin{figure}[tbp]
\begin{center}
\includegraphics[height=5cm]{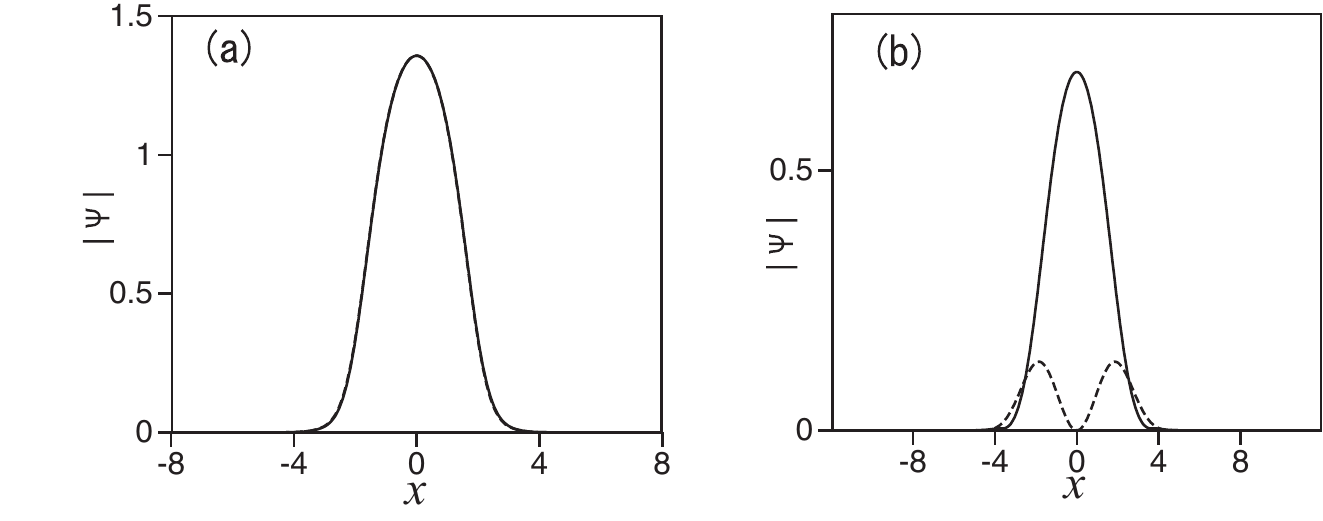}
\end{center}
\caption{(a) Profiles of components $\left\vert \Psi _{+}\right\vert
=\left\vert \Psi _{-}\right\vert $ in a typical stable mixed mode, produced
by Eqs. (\protect\ref{+}) and (\protect\ref{-}) at $\protect\eta =0.1$, $%
\protect\varepsilon =0.4$, $\protect\beta =0.3$, and $\Omega =0$. (b) The
same for a typical semi-vortex, with vorticities $0$ and $2$ in components $%
\Psi _{+}$ and $\Psi _{-}$, respectively, which is produced by the numerical
solution of Eqs. (\protect\ref{+}) and (\protect\ref{-}) in the presence of
the Zeeman splitting with $\Omega =0.2$. Other parameters are $\protect\eta %
=0.4$, $\protect\varepsilon =0.6$, and $\protect\beta =0.6$.}
\label{f3}
\end{figure}

The next step is to consider Eqs. (\ref{+}) and (\ref{-}) in which nonlinear
terms are kept, but the dissipative ones are still ignored:
\begin{gather}
i\partial _{t}\Psi _{+}=-{\frac{1}{2}}(\partial _{x}^{2}+\partial
_{y}^{2})\Psi _{+}+\frac{1}{2}\left( x^{2}+y^{2}\right) \Psi _{+}+(|\Psi
_{+}|^{2}+\alpha |\Psi _{-}|^{2})\Psi _{+}+\beta (\partial _{x}-i\partial
_{y})^{2}\Psi _{-},  \notag \\
i\partial _{t}\Psi _{-}=-{\frac{1}{2}}(\partial _{x}^{2}+\partial
_{y}^{2})\Psi _{-}+\frac{1}{2}\left( x^{2}+y^{2}\right) \Psi _{-}+(|\Psi
_{-}|^{2}+\alpha |\Psi _{+}|^{2})\Psi _{-}+\beta (\partial _{x}+i\partial
_{y})^{2}\Psi _{+}.  \label{nonlin}
\end{gather}%
Solving these equations by means of the imaginary-time method makes it
possible to produce stationary solutions with a fixed norm,%
\begin{equation}
N\equiv \int \int \left[ \left\vert \Psi _{+}\left( x,y\right) \right\vert
^{2}+\left\vert \Psi _{-}\left( x,y\right) \right\vert ^{2}\right] dxdy=1,
\label{N}
\end{equation}%
and eventually identify nonlinear ground states, which realize the minimum
of the Hamiltonian,%
\begin{gather}
H=\int \int \left\{ \frac{1}{2}\left( |\nabla \Psi _{+}|^{2}+|\nabla \Psi
_{-}|^{2}\right) +\frac{1}{2}\left( x^{2}+y^{2}\right) \left( \left\vert
\Psi _{+}\right\vert ^{2}+\left\vert \Psi _{-}\right\vert ^{2}\right)
+\alpha |\Psi _{+}|^{2}\left\vert \Psi _{-}\right\vert ^{2}\right.   \notag
\\
\left. +\frac{1}{2}\left( |\Psi _{+}|^{4}+\left\vert \Psi _{-}\right\vert
^{4}\right) -\beta \left[ \left( (\partial _{x}+i\partial _{y})\Psi
_{+}\right) ^{\ast }(\partial _{x}-i\partial _{y})\Psi _{-}+\mathrm{c.c.}%
\right] \right\} dxdy  \label{E}
\end{gather}%
(here c.c. stands for the complex-conjugate expression), for the given norm.
Figure \ref{f2}(a) shows the so found profiles of $|\Psi _{+}|$ and $|\Psi
_{-}|$ for the ground state at $\beta =0.48,0.45,0.42$. In agreement with
the above-mentioned transition of the ground state from the mixed mode to
the symmetric vortex-antivortex in the linearized system, as the SOC
strength changes from $\beta <0.46$ to $\beta >0.46$, Fig. \ref{f2}
demonstrates that, in the dissipation-free nonlinear system, the same
transition takes place as the SOC strength increases from $\beta =0.45$ to $%
\beta =0.48$.

\subsection{Stable modes in the full nonlinear dissipative system}

The systematic numerical analysis demonstrates that the full system of Eqs. (%
\ref{+}) and (\ref{-}) gives rise to stable stationary states \emph{only if}
the diffusion term is present, $\eta >0$ (at $\eta =0$, chaotic solutions
appear, which are not shown here in detail). For $\beta $ not two large, the
system with small values of $\eta >0$ generates stable patterns of the
mixed-mode type, which thus plays the role of the fundamental state in the
full system, see an example in Fig. \ref{f3}(a).

\begin{figure}[tbp]
\begin{center}
\includegraphics[height=4.7cm]{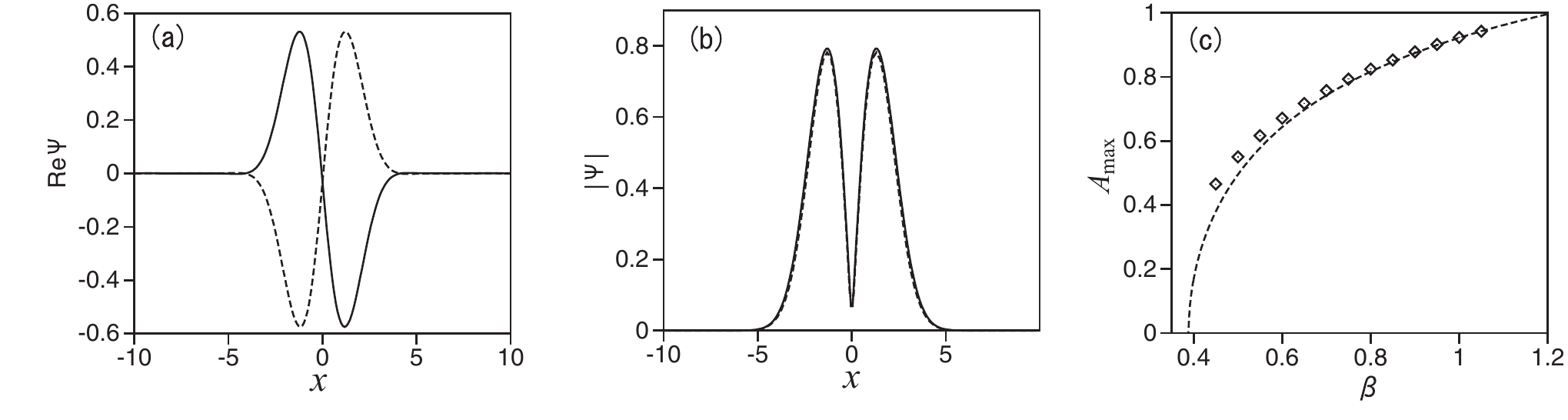}
\end{center}
\caption{(a) Real parts of components $\protect\psi _{+}$ and $\protect\psi %
_{-}$ (solid and dashed lines, respectively) of a stable vortex-antivortex\
produced by the full system at $\protect\varepsilon =0.6,\protect\eta =0.8,$
$\protect\sigma =0.3$, $\protect\beta =0.75$, and $\Omega =0$. (b) The
comparison of $|\protect\psi _{+}|=\left\vert \protect\psi _{-}\right\vert $%
, for the same numerically found vortex-antivortex (solid curve), and the
respective ansatz (\protect\ref{ans}) with $A_{\max }=0.9936$ and $\protect%
\rho =0.5942$ (dashed curve), as predicted by the analytical approximation.
(c) Maximum value $A_{\max }$ of the numerically found $|\protect\psi _{+}|$
component (rhombi) and the respective analytical prediction (the dashed
line) versus the SOC\ strength $\protect\beta $, for the same parameters as
in (a).}
\label{f4}
\end{figure}

At larger values of the SOC strength, $\beta $, and $\eta >0$, the full
system readily produces the fundamental state in the form of stable
vortex-antivortex complexes with the opposite sign of the components, which,
in the linear limit, correspond to the lower sign (antisymmetric bound
state) in Eq. (\ref{vortex-antivortex}). A typical example is presented in
Fig. \ref{f4}, where panel (a) shows real parts of $\Psi _{+}$ and $\Psi _{-}
$ (with the opposite signs, as said here), and panel (b) shows $|\Psi _{\pm
}(x)|$ (the solid line) in the cross section of $y=0$. We stress that the
numerical solution of the full system, including the nonlinear and
dissipative terms, has not produced any stable vortex-antivortex complex
with identical signs of its two component, in spite of the fact that such
complexes play the role of the ground state in the dissipation-free system
close to $\beta =0.5$, as shown above.

The vortex-antivortex states produced by the full system can be quite
efficiently approximated in an analytical form. To this end, the following
ansatz is adopted for their two components:
\begin{equation}
\psi _{\pm }=\pm A_{\max }r\exp \left[ -(1/2)\rho r^{2}\mp i\theta \right] ,
\label{ans}
\end{equation}%
with free parameters $A_{\max }$ and $\rho $, cf. Eq. (\ref%
{vortex-antivortex}). First, ignoring the dissipative terms, it is possible
to predict values of $A_{\max }$ and $\rho $ by minimizing the system's
energy (\ref{E}) at a fixed value of norm $N$ (this time, we do not set $N=1$%
). The results is
\begin{equation}
\rho =\left[ 1+2\beta +(1+\alpha )N/8\right] ^{-1/2},~A_{\max }=\sqrt{N}\rho
.  \label{rho}
\end{equation}%
Further, restoring the dissipative terms $\sim $ $\varepsilon ,\eta $, and $%
\sigma $, the norm, which appears as a free parameter in Eq. (\ref{rho}),
can be predicted by the balance equation (\ref{balance}). The final result is%
\begin{equation}
N=\frac{4}{\sigma ^{2}}\left\{ -\left( \sigma \eta -\epsilon ^{2}(1+\alpha
)/4\right) +\sqrt{\left[ \epsilon ^{2}(1+\alpha )/4-\sigma \eta \right]
^{2}-\sigma ^{2}\left[ \eta ^{2}-\epsilon ^{2}(1+2\beta )\right] }\right\} .
\label{NN}
\end{equation}%
The comparison of the vortex-antivortex modes predicted by Eqs. (\ref{ans})-(%
\ref{NN}) with their numerically found counterparts is presented in Figs. %
\ref{f4}(b) and (c), showing good accuracy of the analytical approximation.
Note that Fig. \ref{f4}(c) demonstrates that this family of the
vortex-antivortex states does not extend to $\beta <0.4$.
\begin{figure}[tbp]
\begin{center}
\includegraphics[height=6cm]{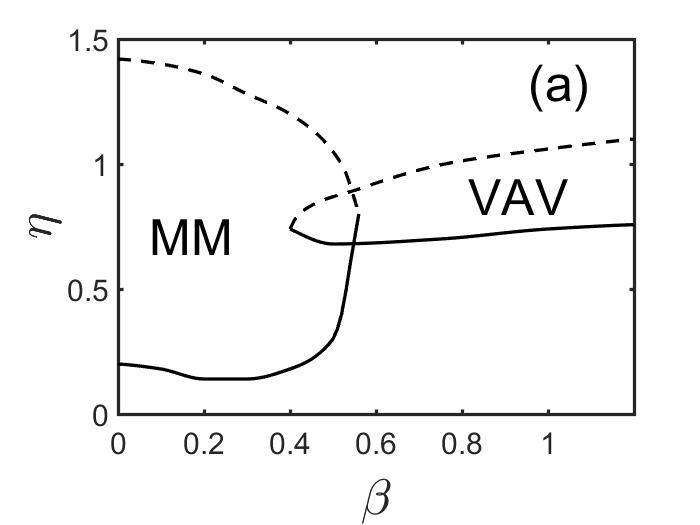} \includegraphics[height=6cm]{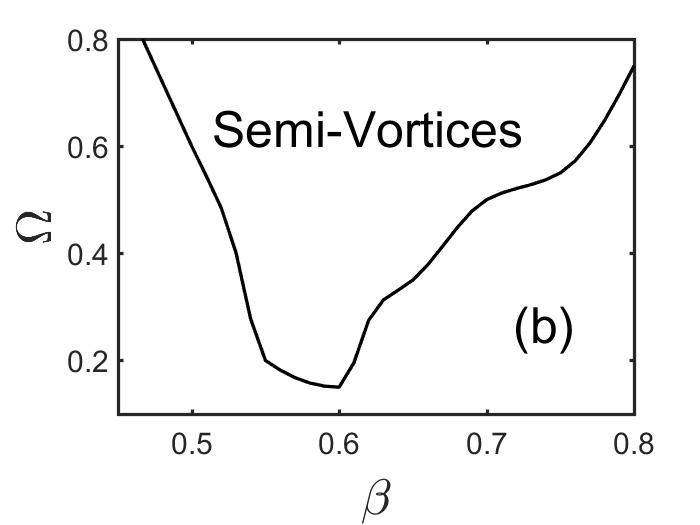}
\end{center}
\caption{(a) Boundaries of areas of the existence of stable mixed mode and
vortex-antivortex states, in the plane of ($\protect\beta ,\protect\eta $).
Other parameters are $\protect\varepsilon =0.6,$ $\protect\sigma =0.3$, and $%
\Omega =0$. (b) The lower existence boundary for semi-vortices in the plane
of $\left( \protect\beta ,\Omega \right) $. Other parameters are $\protect%
\varepsilon =0.6$, $\protect\eta =0.4$, and $\protect\sigma =0.3$.}
\label{f5}
\end{figure}

Results of the systematic numerical analysis are summarized in Fig. \ref{f5}%
(a), which displays the existence areas (between the solid and dashed lines)
for stable mixed-mode and vortex-antivortex states, in the plane of the most
essential control parameters, \textit{viz}., the SOC strength, $\beta $, and
diffusion coefficient, $\eta $ . As mentioned above, the mixed modes and
antisymmetric vortex-antivortex complexes  tend to be stable, as ground
states, at smaller and larger values of $\beta ,$ respectively, with a small
bistability region revealed by the figure. Above the dashed lines, the
solutions decay to zero, while below the solid lines, the evolution makes
them chaotic.

\section{Stable semi-vortex states supported \ by the Zeeman splitting}

The Zeeman splitting, represented by $\Omega \neq 0$ in Eqs. (\ref{+}) and (%
\ref{-}), can strongly impact properties of nonlinear modes in the present
system, as is known to happen in other polariton models, see, e.g., Ref.
\cite{gul}. We here aim to outline general trends imposed by the Zeeman
splitting, without producing full details. As suggested by the analysis of
the spin-orbit-coupled two-component atomic condensates \cite{Sherman}, the
enhancement of the Zeeman splitting tends to replace mixed modes by the
semi-vortices, in which, in the present case, one component has zero
vorticity, and the other carries vorticity $2$. The present system does not
give rise to semi-vortices  at $\Omega =0$; however, they appear with the
increase of $\Omega $.

The case of very large $\Omega $ can be considered by means of an analytical
approximation. In this case, the chemical potential contains a large term, $%
-\Omega $, suggesting one to substitute
\begin{equation}
\Psi _{\pm }\left( x,y,t\right) \equiv e^{i\Omega t}\Phi _{\pm }\left(
x,y,t\right) ,  \label{Omega}
\end{equation}%
where the time dependence in $\Phi _{\pm }$ is assumed to be slow in
comparison to $e^{i\Omega t}$. Then, using an approximation similar to that
developed in Ref. \cite{Sherman}, we use Eq. (\ref{+}) to eliminate $\Phi
_{+}$ in favor of $\Phi _{\_}$:
\begin{equation}
\Phi _{+}\left( x,y,t\right) \approx -\frac{\beta }{2\Omega }(\partial
_{x}-i\partial _{y})^{2}\Phi _{-}\left( x,y,t\right) ~.  \label{PsiPsi}
\end{equation}%
Finally, the substitution of this approximation in Eq. (\ref{-}) reduces it
to a single\textit{\ fourth-order} equation for the wave function $\Phi _{-}$%
:
\begin{eqnarray}
&&i\partial _{t}\Phi _{-}=-{\frac{1}{2}}\left( 1-i\eta \right) (\partial
_{x}^{2}+\partial _{y}^{2})\Phi _{-}+|\Phi _{-}|^{2}\Phi _{-}-\frac{\beta
^{2}}{2\Omega }(\partial _{x}^{2}+\partial _{y}^{2})^{2}\Phi _{-}  \notag \\
&&+i(\varepsilon -\sigma |\Phi _{-}|^{2})\Phi _{-}+\frac{1}{2}%
(x^{2}+y^{2})\Phi _{-}.  \label{Psi}
\end{eqnarray}%
In particular, in the case of $\eta =\varepsilon =\sigma =0$, we arrive at a
linear or nonlinear Schr\"{o}dinger equation with a kinetic \textit{%
super-energy} term, expressed by the fourth-order operator $\sim \beta
^{2}/\Omega $:%
\begin{equation}
i\partial _{t}\Phi _{-}=-{\frac{1}{2}}(\partial _{x}^{2}+\partial
_{y}^{2})\Phi _{-}-\frac{\beta ^{2}}{2\Omega }(\partial _{x}^{2}+\partial
_{y}^{2})^{2}\Phi _{-}+|\Phi _{-}|^{2}\Phi _{-}+\frac{1}{2}(x^{2}+y^{2})\Phi
_{-}.  \label{super}
\end{equation}%
It is obvious that the present approximation gives rise to semi-vortices,
with a large zero-vorticity component produced by isotropic equation (\ref%
{Psi}), and a small vortex one generated from it by Eq. (\ref{PsiPsi})

An example of a stable semi-vortex is presented in Fig. \ref{f3}(b), which
shows profiles of absolute values of the zero-vorticity component (the solid
curve), $\left\vert \Psi _{+}\right\vert $, and its counterpart with
vorticity $2$ ($|\Psi _{-}|$, the dashed line). Results of the systematic
study of the semi-vortices  are summarized in Fig. \ref{f5}(b), in the
parameter plane of $(\beta ,\Omega )$. Stable semi-vortices  exist above the
boundary shown in the figure, provided that $\Omega $ exceeds a finite
threshold value, $\Omega _{\min }\approx 0.16$.

\section{Summary}

The aim of this work is to identify species of robust 2D localized modes
which play the role of fundamental states in two-component (spinor)
exciton-polariton condensates, subject to the action of SOC (spin-orbit
coupling), represented by the second-order linear differential operator
which mixes the two components. The system includes ordinary terms which are
common to semiconductor-microcavity models, such as the linear gain and
effective diffusion, nonlinear loss, self-repulsive nonlinearity in each
component, and the isotropic harmonic-oscillator trapping potential.
Starting from the analysis of the linearized dissipation-free system, and
then proceeding to its full form, we have found, by means of numerical and
analytical methods, that basic states supported by the system are
antisymmetric vortex-antivortex complexes and mixed modes. The latter ones
combine zero-vorticity and vorticity-carrying terms in each component. The
mixed modes and vortex-antivortex complexes tend to realize the fundamental
states (similar to the ground states of conservative systems) when SOC is,
respectively, weak or strong, with a small region of bistability. The
presence of the effective diffusion is a necessary condition for the
stability of modes of both types in the full system. We have also found that
the addition of the Zeeman splitting tends to replace these modes by stable
semi-vortices, with vorticities $0$ in the larger component and $2$ in the
smaller one.

\section*{Funding}

Royal Society (IE 160465); ITMO University through the Government of Russia
(074-U01); Binational (US-Israel) Science Foundation (2015616).

\end{document}